%% file: cranelift.tex
\documentclass[acmsmall,screen]{acmart}
\AtBeginDocument{%
  \providecommand\BibTeX{{%
    \normalfont B\kern-0.5em{\scshape i\kern-0.25em b}\kern-0.8em\TeX}}}

\usepackage{graphicx} 
\usepackage{stfloats} 
\usepackage{enumitem}
\usepackage{multirow}
\usepackage{longtable}
\usepackage[most]{tcolorbox}
\usepackage{tabu}
\usepackage{listings}
\usepackage{url}
\usepackage{amsmath}
\usepackage{amsfonts}
\usepackage{mathtools}
\usepackage{amsthm}
\usepackage{threeparttable}
\usepackage{caption}
\usepackage{color}
\usepackage{array}
\usepackage{wasysym}
\usepackage{algorithm}  
\usepackage{algorithmicx}  
\usepackage[noend]{algpseudocode}
\usepackage{makecell}
\usepackage{subcaption}
\usepackage{xcolor}
\usepackage[normalem]{ulem}
\usepackage[framemethod=TikZ]{mdframed} 
\usepackage{xcolor} 

\usepackage{wrapfig}
\usepackage{soul}

\mdfdefinestyle{mygraybox}{
    backgroundcolor=gray!20,  
    roundcorner=8pt,          
    linewidth=0pt,            
    innertopmargin=6pt,       
    innerbottommargin=6pt,    
    innerleftmargin=8pt,      
    innerrightmargin=8pt      
}

\author{Shangtong Cao}
\affiliation{%
 \institution{Beijing University of Posts and Telecommunications}
 \country{China}
 }

\author{Tianlei Song}
\affiliation{%
 \institution{Harbin Institute of Technology, Shenzhen}
 \country{China}
}

\author{Qiuping Yi}
\affiliation{%
 \institution{Beijing University of Posts and Telecommunications}
 \country{China}
}

\author{Tianyu Chen}
\affiliation{%
 \institution{Peking University}
 \country{China}
}

\author{Guoai Xu}
\affiliation{%
 \institution{Harbin Institute of Technology, Shenzhen}
 \country{China}
}

\author{Ningyu He}
\affiliation{%
 \institution{The Hong Kong Polytechnic University}
 \country{China}
}

\author{Haoyu Wang*}
\affiliation{%
 \institution{Huazhong University of Science and Technology}
 \country{China}
}

\definecolor{lightgreen}{rgb}{0.25, 0.63, 0.4375}
\definecolor{darkblue}{rgb}{0.02, 0.16, 0.49}

\definecolor{darkgreen}{rgb}{0, 0.5, 0}
\definecolor{darkred}{rgb}{0.72,0.04,0.04}

\newcommand{\code}[1]{{\ttfamily #1}}
\definecolor{commentcolor}{RGB}{31,135,29}

\lstdefinestyle{custom}{
  commentstyle=\color{commentcolor},
  moredelim=[il][\textcolor{commentcolor}]{\#}
}

\setcopyright{none}
\renewcommand\footnotetextcopyrightpermission[1]{}

\newcommand{\framework}{\textsc{CLIR}}
\begin{document}

\title{{\framework}: Liveness-Driven and Structure-Aware Fuzzing for the Cranelift Compiler}

\renewcommand{\shortauthors}{Trovato et al.}

\begin{abstract}
Modern compilers are complex software systems that must correctly translate high-level programming languages into machine code across multiple architectures. Cranelift, a fast and modern compiler backend originally developed for WebAssembly and recently adopted as an experimental backend for Rust, has gained increasing importance due to its superior compilation speed compared to LLVM and comprehensive multi-architecture support, including x86-64, AArch64, s390x, and RISCV64. However, despite decades of development in compiler testing, testing Cranelift still presents unique challenges, including (1) constructing valid IR under the strict enforcement of SSA form, (2) generating sequences with sufficient computational density to stress backend components, and (3) balancing broad backend coverage with efficient root cause analysis across heterogeneous architectures.

To address these challenges, we propose {\framework}, a differential testing framework that integrates a syntax-preserving hierarchical generation strategy to guarantee SSA validity, a liveness-guided instruction refinement mechanism to maximize computational density, and a diagnosis-guided cross-architecture adaptation scheme to facilitate efficient root cause analysis across heterogeneous backends. 
Our comprehensive evaluation demonstrates that {\framework} significantly outperforms existing state-of-the-art baselines, detecting 8$\times$, 24$\times$, and 8$\times$ more unique bugs than cranelift-fuzzgen, wasm-smith, and WASMaker, respectively, while RustSmith uncovered no bugs.
Consequently, within 72 hours of testing, {\framework} discovered 24 bugs spanning all target architectures, with 21 confirmed and 9 fixed. 
\end{abstract}

\maketitle

\input{sec-1-intro.tex}
\input{sec-2-background.tex}

\input{sec-4-approach.tex}

\input{sec-5-implement-and-evaluation.tex}
\input{sec-6-discussion}
\input{sec-7-related-work}

\bibliographystyle{ACM-Reference-Format}
\bibliography{cranelift}

\end{document}

%% file: sec-1-intro.tex
\section{Introduction}
\label{sec:intro}
As sophisticated translation systems, modern compilers bridge the gap between high-level programming languages and diverse machine architectures.
Despite decades of development and testing, even well-established compilers continue to harbor numerous bugs. 
For instance, LLVM~\cite{llvm}, one of the most widely used compiler infrastructures, has accumulated over 26,000 reported issues throughout its development history, with more than 3,000 new bugs discovered annually~\cite{llvmissues}. 
GCC has accumulated tens of thousands of reported bugs over its lifetime~\cite{gcc,sun2016toward}, demonstrating that \textit{compiler correctness remains a persistent challenge for mature compiler systems}.

The goal of Cranelift is to be a fast, safe, and modern compiler, originally developed as the JIT and AOT backend for the Wasmtime WebAssembly runtime~\cite{cranelift}, while currently it has been adopted as an experimental backend for the Rust compiler due to its significantly faster compilation speed compared to LLVM~\cite{fast}. 
With Rust becoming a popular choice for developing low-level systems~\cite{rust1,rust2,rust3,rust4}, \textit{the correctness of Cranelift becomes the role under the spotlight}.
Consequently, the growing adoption of Cranelift in production environments, coupled with its relative novelty compared to GCC and LLVM, makes comprehensive testing for it an \textit{inevitable} topic.

One straightforward approach to testing Cranelift is to leverage existing ecosystem tools, such as RustSmith~\cite{rustsmith} or wasm-smith~\cite{wasm-smith}, to generate high-level source code and compile it down to IR. However, \textit{this top-down strategy is insufficient due to the \textit{translation gap}}, \textit{i.e.,} the frontend compilation process inherently normalizes and sanitizes inputs, filtering out many valid but edge-case IR patterns~\cite{irfuzzer}. Consequently, a wide range of legal IR instruction sequences, which do not correspond to standard high-level source constructs, remain unreachable and untested. 
Furthermore, as a general-purpose backend intended to support diverse frontends, Cranelift must be robust against arbitrary valid IR inputs, not merely those produced by specific languages. The necessity of direct IR-level testing is explicitly validated by the Cranelift community through the maintenance of \texttt{cranelift-fuzzgen}~\cite{fuzzgen}, an official tool dedicated to raw IR generation.

However, \textit{performing testing against Cranelift is still challenging}, where three major challenges must be addressed. 
First, constructing syntactically valid IR is non-trivial due to Cranelift's strict SSA constraints. The generator must ensure that every variable use is dominated by its definition across complex control flows to avoid immediate rejection before the compilation commences. 
Second, generated IR must exhibit high computational complexity to be effective. Naively generated code often lacks deep data dependencies and is susceptible to trivial simplification. To this end, the challenge lies in systematically synthesizing instructions with sufficient computational density and interconnectivity to stress critical backend components. 
Third, adapting to heterogeneous backends introduces a dilemma between compatibility and coverage. The framework must support diverse hardware architectural constraints while providing automated mechanisms to efficiently pinpoint the root causes of bugs amid massive failure reports.

\textbf{This work.} 
In this work, we propose {\framework}, a comprehensive Cranelift compiler testing framework designed to bridge the gap between syntactic validity and testing utility. To achieve this, we design a \textit{Syntax-Preserving Hierarchical SSA-form IR Generation} strategy that constructs test cases in a top-down manner. This approach extracts basic blocks from real-world Rust and WebAssembly programs and assembles them via a dominator-driven algorithm to strictly enforce SSA compliance. Based upon this, {\framework} further applies a \textit{Liveness-Guided Instruction Refinement} mechanism to maximize testing utility. By systematically anchoring data dependency chains to observable behaviors and coupling block parameters, this strategy prevents trivial simplification and ensures the generated code exhibits high computational complexity. To efficiently investigate inconsistent behaviors across heterogeneous backends, {\framework} incorporates a \textit{Diagnosis-Guided Cross-Architecture Adaptation} strategy. This component utilizes dual-mode profiling to tailor test cases for specific architectures and employs hierarchical static instrumentation to pinpoint root causes at the instruction level. 

We conduct extensive experiments to show the superiority of {\framework} over state-of-the-art techniques in terms of both effectiveness and coverage.
Specifically, by applying {\framework} to four target architectures, \textit{i.e.,} x86-64, AArch64, RISCV64, and s390x, over a 72-hour period, we identified 24 bugs, which is 8$\times$, 24$\times$, and 8$\times$ more than those detected by \texttt{cranelift-fuzzgen}, \texttt{wasm-smith}, and \texttt{WASMaker}, respectively, while the high-level fuzzer \texttt{RustSmith} failed to detect any bugs. Moreover, {\framework} achieves 1.2$\times$ higher code coverage across all supported architectures. 
With our timely disclosure, 21 bugs have been confirmed by Cranelift developers, and 9 of them have been fixed by the time of this writing.

\textbf{Our contribution.} We summarize the major contributions of this work as follows:
\begin{itemize}[leftmargin=*]
    \item \textbf{Cranelift IR Generator.} We propose a robust Cranelift IR generator that combines syntax-preserving hierarchical generation with liveness-guided instruction refinement. This approach efficiently produces syntactically valid and computationally complex Cranelift IRs by strictly enforcing SSA constraints while ensuring global data dependencies to maximize testing utility.

    \item \textbf{Impactful Framework.} We implement {\framework}, a Cranelift compiler testing framework featuring diagnosis-guided adaptation. By integrating signature-guided failure clustering and feedback-driven profile adaptation, it efficiently handles heterogeneous backends, automating root cause localization and reducing manual triage efforts.

    \item \textbf{Comprehensive Evaluation.} We conduct a comprehensive evaluation demonstrating that {\framework} significantly outperforms state-of-the-art baselines in both bug-finding capability and generated IR complexity. Within 72 hours, it uncovered 24 unique bugs, with 21 confirmed and 9 fixed, proving its superior effectiveness in exposing deep backend flaws.

    \item \textbf{Open Source.} We open source {\framework} and our dataset at \href{https://github.com/CLIR479/CLIR}{link} to facilitate future research in Cranelift compiler testing.
\end{itemize}

%% file: sec-2-background.tex
\section{Background}
\label{sec:background}
In this section, we illustrate necessary background about Cranelift and the corresponding Cranelift intermediate representation (IR).

\subsection{Cranelift Compiler}
\label{sec:backgroud:cranelift}
Cranelift is a code generator designed to be a fast, secure, and relatively simple compiler backend, implemented entirely in Rust. Unlike traditional backends, like LLVM, which prioritize peak execution performance often at the cost of compilation time, Cranelift focuses on achieving high compilation speed and low memory footprint, making it particularly suitable for Just-In-Time (JIT) compilation scenarios~\cite{jit}. 
It accepts a target-independent Intermediate Representation (Cranelift IR) as input and translates it into executable machine code for various architectures, including x86-64, AArch64, s390x~\cite{s390x}, and RISCV64~\cite{riscv64}. 
While originally developed as the JIT and AOT engine for the Wasmtime virtual machine~\cite{wasmtime} to execute WebAssembly, its general-purpose design has led to its adoption as an experimental backend for the Rust compiler~\cite{cranelift-rust}, offering significantly faster debug build times.

The core transformation from platform-independent IR to target-specific machine instructions, \textit{i.e.,} \textit{lowering}, is driven by a domain-specific language called ISLE (Instruction Selection Lowering Expressions)~\cite{isle}. 
Instead of hard-coding instruction selection logic, Cranelift defines these rules declaratively within \texttt{.isle} files for each architecture. These files consist of term-rewriting rules that explicitly map patterns in the Cranelift IR to their corresponding machine instructions. 
Consequently, the \texttt{.isle} files serve as the central repository for backend-specific lowering logic, making them a critical component in ensuring the correctness of code generation.

\subsection{Cranelift IR Instance}
\label{sec:backgroud:ir instance}
Cranelift uses a strongly typed intermediate representation in Static Single-assignment (SSA) form~\cite{ssa}, where each value is assigned exactly once and is immutable thereafter. 
Fig.~\ref{fig:background:cranelift ir} illustrates a concrete example of a C function compiled to Cranelift. 
Syntactically, a function definition begins with a signature (L1\footnote{L1 refers to the first line, we adopt such notations in the following.}), specifying the function name (\texttt{\%sum}), input parameter types, and return types. 
The body consists of a sequence of \textit{Extended Basic Blocks} (EBBs), starting with the entry block (\texttt{block0}).
Several structural distinctions exist between Cranelift IR and other ones like LLVM:

\begin{figure}[t] 
    \centering
    \includegraphics[width=\columnwidth]{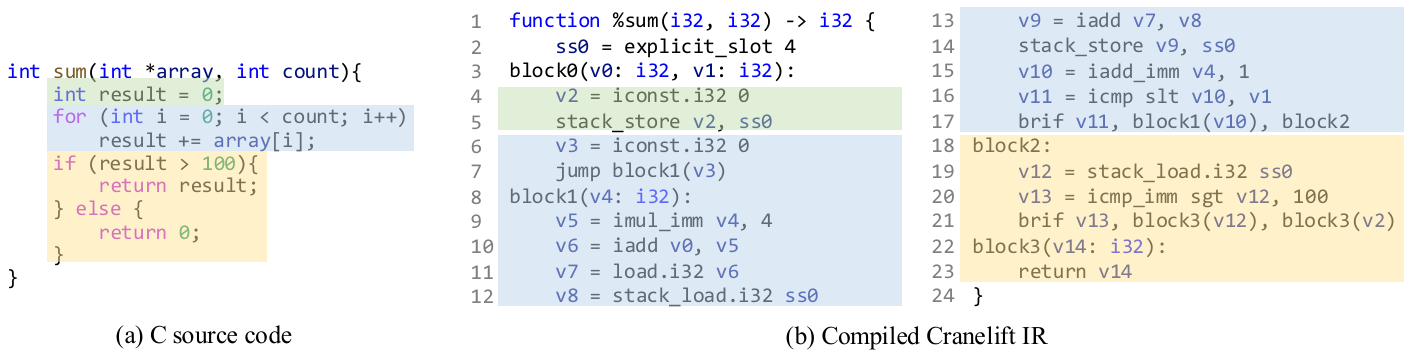} 
    \vspace{-0.2in}
    \caption{A C code snippet and the corresponding Cranelift IR program.}
    \vspace{-0.1in}
    \label{fig:background:cranelift ir} 
\end{figure}

\textit{Function Preamble.}
Before the first basic block, Cranelift employs a \textit{function preamble} to declare various entities required by the function body, like stack slots (L2).
Unlike LLVM, which typically uses inline instructions like \texttt{alloca} for allocation, Cranelift requires these entities to be defined upfront.
In Fig.~\ref{fig:background:cranelift ir}, the preamble explicitly defines a stack slot, which is subsequently accessed via specific instructions such as \texttt{stack\_store} (L5) and \texttt{stack\_load} (L12), ensuring precise control over resource usage.

\textit{Block Parameters.}
The most significant feature of Cranelift IR is its handling of SSA construction. 
Instead of using \textit{Phi nodes}~\cite{phi} to merge values from different control flow predecessors, Cranelift employs \textit{block parameters}. 
Each basic block can explicitly declare typed parameters, acting similarly to function arguments.
For instance, the loop header \texttt{block1} declares a parameter \texttt{v4: i32} (L8).
The value of \texttt{v4} is determined by the arguments passed during the jump to this block:
\begin{itemize}[leftmargin=*]
    \item From the entry block (L7), the instruction \texttt{jump block1(v3)} passes the initial value \texttt{v3} to \texttt{v4}.
    \item From the loop back-edge (L17), the instruction \texttt{brif} passes the incremented value \texttt{v10} to \texttt{v4}.
\end{itemize}
This design unifies control flow transfer with data flow, simplifying the IR structure and making it easier to analyze and generate programmatically.

\textit{Instruction Syntax.}
Instructions in Cranelift follow a three-address code format, typically consisting of an \textit{opcode}, \textit{type suffixes}, and \textit{operands} (\textit{e.g.,} \texttt{v5 = imul\_imm v4, 4} in L9). 
Type suffixes (like \texttt{.i32} in \texttt{iconst.i32}) are mandatory for polymorphic instructions to ensure strict type safety.
Terminator instructions, such as \texttt{brif} (conditional branch) and \texttt{jump} (unconditional branch), are strictly required at the end of every block to explicitly define the control flow graph.

\section{Challenges}
\label{sec:challenge}
Compiler testing is challenging. Despite being a well-studied topic, modern compilers like GCC and LLVM continue to exhibit numerous bugs, with historical reports showing they have accumulated tens of thousands of issues throughout their development~\cite{gcc,llvmissues}. 
While traditional compilers have been extensively tested, newer compiler infrastructures, such as Cranelift, remain less explored.
Generally, comprehensive compiler testing relies on two vital components: the \textit{seed} (test case) and the \textit{oracle}. Taking advantage of differential testing allows us to bypass the oracle problem, shifting the burden entirely to the quality of the generated seeds.
However, testing Cranelift presents unique hurdles due to its specific IR design and multi-backend nature.
We summarize the three major challenges as follows:

\noindent 
\textbf{Challenge \#1: Constructing syntactically valid IR under strict SSA constraints}. The primary prerequisite for backend testing is ensuring syntactic validity, which, however, \ul{is inherently difficult because the generator must satisfy rigorous dominance and type constraints while maintaining the structural diversity required for effective fuzzing.} Cranelift IR enforces a strongly typed SSA form (\S\ref{sec:backgroud:ir instance}), introducing rigorous dependencies where every variable usage must be dominated by its definition. 
In the presence of complex control flow containing loops and branches, maintaining these dominance invariants is non-trivial. While directly compiling high-level languages (\textit{e.g.,} Rust or Wasm) to obtain Cranelift IR ensures validity, such approaches suffer from \textit{frontend shielding} (\S\ref{sec:intro}), which canonicalizes input and obscures backend-specific bugs. 
Consequently, direct IR generation is required, yet existing methods falter, \textit{i.e.,} \textit{naive fuzzing}~\cite{afl} lacks structural awareness, leading to frequent dominance violations, while \textit{grammar-based generation}~\cite{csmith} relies on computationally expensive constraint solving to track global variable scoping.

\noindent
\textbf{Challenge \#2: Generating Computationally Complex IR}.
Merely generating syntactically valid IR is insufficient, as \ul{effective test cases must also exhibit high computational complexity to stress critical compiler backend components} like register allocation and instruction scheduling.
Prior works like IRFuzzer~\cite{irfuzzer} and Alive-mutate~\cite{fan2024high} rely on stochastic generation or local mutation.
However, these approaches lack guidance to construct deep data dependencies. Also, their generation is hampered by uniform randomness, while mutation is bounded by the original seed complexity.
Therefore, the challenge lies in systematically constructing IR that possesses sufficient computational density and interconnectivity to avoid trivial simplification, without compromising the strict correctness guarantees established in \textbf{Challenge \#1}.

\noindent
\textbf{Challenge \#3: Adapting to heterogeneous backends and efficient bug diagnosis}.
Cranelift is designed as a cross-platform code generator supporting diverse architectures (\textit{e.g.,} x86-64, AArch64, RISCV64, s390x). This introduces a dilemma between \textit{compatibility} and \textit{coverage}.
On one hand, different backends have distinct constraints, like different instruction sets (ISAs), specific operand types, and unique calling conventions. A generator must be architecture-aware to avoid producing IR that is valid generally but unsupported on a specific target, while simply generating compatible intersection sets limits the ability to test backend-specific features.
On the other hand, in the context of differential testing, a single underlying compiler bug often manifests as thousands of failing test cases across different seeds. Manually triaging these failures is impractical.
Thus, the final challenge is to design a framework that can \ul{flexibly adapt to heterogeneous backend constraints while providing automated mechanisms to pinpoint bug root causes}, ensuring testing efficiency.

To tackle \textbf{Challenge \#1}, we propose a \textit{syntax-preserving hierarchical SSA-form IR generation} approach that strictly adheres to SSA constraints and type consistency (\S\ref{sec:method:m1}). 
To address \textbf{Challenge \#2}, we introduce a \textit{liveness-guided instruction refinement} strategy that maximizes computational complexity, thereby ensuring the generated IR possesses high testing utility for stressing backend components (\S\ref{sec:method:m2}).
As for \textbf{Challenge \#3}, we design a \textit{diagnosis-guided cross-architecture adaptation} mechanism to guide computing resources on identifying bugs with different root causes (\S\ref{sec:method:m3}).

%% file: sec-4-approach.tex
\section{Methodology} 
\label{sec:approach}

\begin{figure*}[t] 
    \centering 
    \includegraphics[width=\linewidth]{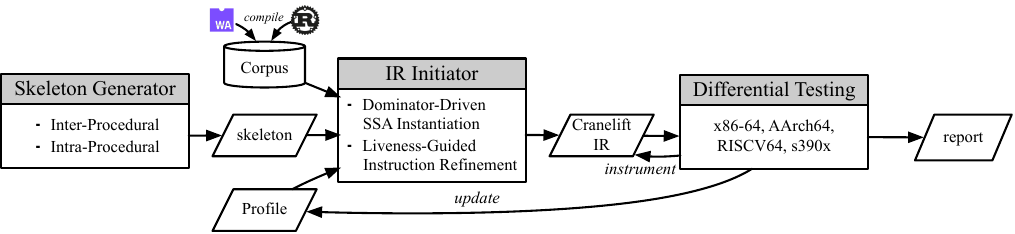} 
    \vspace{-0.2in}
    \caption{The workflow of {\framework}.} 
    \vspace{-0.2in}
    \label{fig:workflow} 
\end{figure*}

\subsection{Overview}

As shown in Fig~\ref{fig:workflow}, the workflow of {\framework} can be divided into three modules: \textit{skeleton generator}, \textit{IR Initiator}, and \textit{Differential Testing}.
First, the \textit{skeleton generator} establishes the structural foundation of the test case by constructing an inter-procedural call graph and recursively nesting atomic control structures to form complex intra-procedural CFGs, resulting in a function skeleton.
Next, the \textit{IR initiator} populates this skeleton to produce valid Cranelift IR. It draws from a \textit{Corpus} of real-world Cranelift instruction snippets (see \S\ref{sec:method:corpus}) and utilizes a dominator-driven algorithm to ensure SSA compliance and type safety (see \S\ref{sec:method:m1}), while also applying instruction refinement to enhance testing utility (see \S\ref{sec:method:m2}).
Finally, the \textit{differential testing} module executes the generated IR across multiple architectures, \textit{i.e.,} x86-64, AArch64, RISCV64, and s390x. This phase incorporates an instrumentation loop for root cause localization and a feedback-driven adaptation that refines the architecture \textit{profile} to guide future generation (see \S\ref{sec:method:m3}).

\subsection{Cranelift Corpus Preparation}
\label{sec:method:corpus}

We leverage existing representative Rust and WebAssembly programs from crates.io~\cite{crates} and WasmBench benchmark suite~\cite{wasmbench_github} to prepare the corpus.
First, we compile collected Rust and WebAssembly programs into Cranelift IR. Then, we extract all basic blocks and store them as \textit{reusable units} in the corpus. Each basic block is treated as an individual code snippet that captures realistic instruction patterns and operand usages commonly found in real-world programs. 
We underline that \textit{this corpus serves as the cornerstone for the later stages}. Its rich and diverse semantics significantly increase the likelihood of triggering compiler bugs. Such corpus-based synthesizing methods are widely adopted in related testing work~\cite{li2024boosting,zhong2022enriching,wasmaker,jiang2025distinguishability}.

\subsection{Syntax-Preserving Hierarchical SSA-form IR Generation}
\label{sec:method:m1}
{\framework} adopts a hierarchical, top-down strategy to construct IR test cases, ensuring structural validity from the global scope down to individual instructions. The generation process operates across three distinct layers as follows. 
First, it establishes the inter-procedural skeleton by defining function signatures and their invocation relationships (\S\ref{sec:method:generation:function}). 
Second, within each function, it constructs a complex Control Flow Graph (CFG) by recursively nesting atomic control structures (\S\ref{sec:method:generation:CFG}).
Finally, it initiates these basic blocks with concrete instructions using a dominator-driven algorithm to satisfy strict SSA Def-before-Use constraints (\S\ref{sec:method:ssa}).

\subsubsection{Inter-Procedural Skeleton Generation}
\label{sec:method:generation:function}
This stage aims to construct a diverse function call graph.
Starting with a designated entry function, the generator recursively declares and invokes a set of sub-functions. 
To simulate the complexity of real-world software, we enforce diversity in function signatures, \textit{i.e.,} randomizing parameters and return values, which can increase the likelihood of exposing backend bugs related to stack frame management and parameter passing.

Furthermore, we incorporate a wide range of calling mechanisms. Cranelift supports multiple invocation styles including direct calls, indirect calls (function pointers), and tail calls (\textit{e.g.,} \code{Call}, \code{CallIndirect}, \code{ReturnCall}). By integrating these variations into the generation process, {\framework} ensures comprehensive coverage of the compiler's function transition logic.

\begin{figure}[t]
    \centering
    \begin{subfigure}[c]{0.6\columnwidth} 
        \centering
        \includegraphics[width=\linewidth]{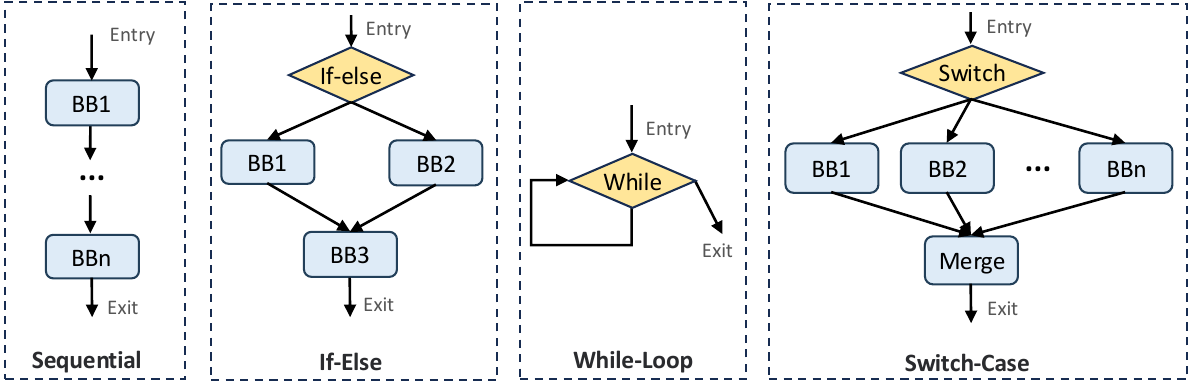}
        \caption{Atomic structures}
        \label{fig:atomic}
    \end{subfigure}%
    \hfill %
    \begin{subfigure}[c]{0.32\columnwidth} 
        \centering
        \includegraphics[width=\linewidth]{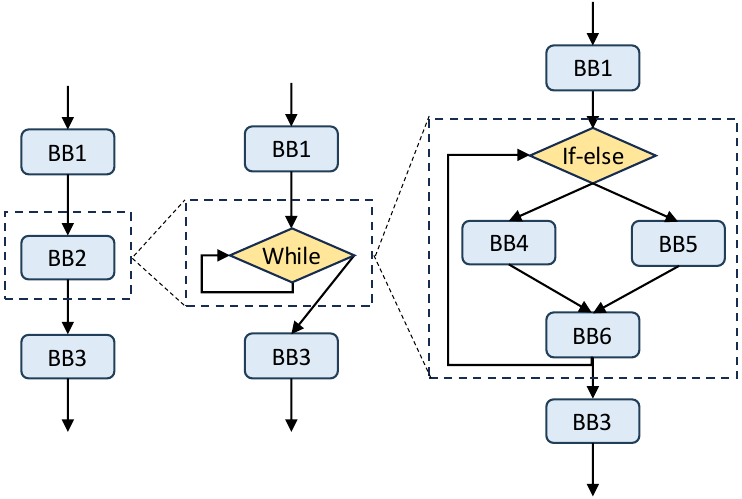}
        \vspace{-0.3in}
        \caption{Construction example}
        \label{fig:example}
    \end{subfigure}
    
    \caption{Overview of Intra-Procedural CFG Construction. (a) The four atomic structures used as building blocks; (b) An example of recursively constructing a CFG using these structures.}
    
    \label{fig:cfg_combined}
\end{figure}

\subsubsection{Intra-Procedural CFG Construction}
\label{sec:method:generation:CFG}
Within each function, constructing a non-trivial CFG is critical as core compiler components, such as dominator tree construction, register allocation, and loop analysis, rely heavily on the topology of the CFG. 
Advanced optimizations, including loop-invariant code motion and branch prediction, are only triggered by specific control flow patterns~\cite{advanced}. 
Obviously, linear or overly simple flows fail to exercise these passes, leaving potential logic errors undetected.

To this end, we propose a \textit{recursive CFG-substructure substitution} method to build complex yet valid CFGs within each function.
Specifically, we define four atomic control flow structures: \textit{Sequential}, \textit{If-Else}, \textit{While-Loop}, and \textit{Switch-Case}, as illustrated in Fig.~\ref{fig:atomic}.
A key property of these atomic structures is that they maintain a \textit{Single-Entry-Single-Exit} structure, \textit{i.e.,} the indegree of the entry block and the outdegree of the exit block are always one.
This property allows us to \textit{treat any atomic structure as functionally equivalent to a single basic block}, allowing us to iteratively replace a basic block within an existing structure with a more complex atomic structure.
For instance, as shown in Fig.~\ref{fig:example}, a block within a \textit{Sequential} structure can be replaced by a \textit{While-Loop} structure, which can subsequently be replaced by an \textit{If-Else} structure.
The substitution process preserves graph integrity by 1) redirecting incoming edges from the original block's predecessors to the new structure's entry, and 2) connecting the new structure's exit to the original block's successors.
This recursive nesting allows {\framework} to generate arbitrarily complex control flows within functions, effectively increasing the structural intricacy while ensuring logical correctness.

\renewcommand{\algorithmicrequire}{\textbf{Input:}}
\renewcommand{\algorithmicensure}{\textbf{Output}}

\begin{algorithm}[t] 
    \caption{Dominator-Driven SSA Instantiation.}
    \label{algorithm:function generation}
    \begin{algorithmic}[1] 
        \Require $s$ - the block-level skeleton; $cps$ - a corpus of Cranelift IR instruction snippets (see \S\ref{sec:method:corpus})
        
        \State $defUseMap \leftarrow \emptyset$
        \State $domMap \leftarrow \operatorname{getDominance}(s)$
        \State $entryBlock \gets \operatorname{getEntry}(s)$
        \State $\operatorname{initBlockParam}(entryBlock)$ 
        \State $\operatorname{setDefUse}(defUseMap, entryBlock)$ 
        
        \For{$block\ \text{in}\ \operatorname{dfs}(s)$}
            \State $dominators \gets domMap[block]$
            \State $instrs \gets \operatorname{sample}(cps)$
            \For{$instr\ \text{in}\ instrs$}
                \If{$\operatorname{random}() > \sigma$}
                    \State $instr \gets \operatorname{mutateOp}(instr)$
                \EndIf
                \State $\operatorname{fillRef}(dominators, defUseMap, instr)$ \Comment{\S\ref{sec:method:m2:prior}}
                \State $\operatorname{setDefUse}(defUseMap, block)$
            \EndFor
        \EndFor
    \end{algorithmic}
\end{algorithm}

\subsubsection{Dominator-Driven SSA Instantiation}
\label{sec:method:ssa}
Following the constructed control flow skeleton, the final phase involves populating the basic blocks with concrete IR instructions.
To address \textbf{Challenge \#1}, we propose a \textit{dominator-based instruction generation} method (see Algorithm~\ref{algorithm:function generation}), which utilizes the dominance analysis to maintain a valid pool of live variables for operand selection, ensuring syntactic correctness by design.

Specifically, as outlined in Algorithm~\ref{algorithm:function generation}, the process accepts the block-level function skeleton generated by \S\ref{sec:method:generation:CFG} and a corpus of Cranelift IR instruction snippets (detailed in \S\ref{sec:method:corpus}) as inputs.
First, we initialize $defUseMap$ to maintain the def-use relationships of variables, domination relationships among blocks $domMap$ extracted by the skeleton, and the entry block (L1--L3).
At L4, for the entry block, we explicitly add block parameters (like L3 and L8 in Fig.~\ref{fig:background:cranelift ir}) to create \textit{root nodes} for all operand types (\textit{e.g.,} \texttt{i32} and \texttt{i8x16}) in case the following instructions need a reference to such data types. L5 updates $defUseMap$ to record that these variables are \textit{defined} in the entry block.
We then traverse the skeleton in a DFS manner (L6). For each block, we first extract its control-flow dominators (L7), and sample an instruction snippet from the corpus to try to initiate this block (L8).
Instead of directly copy-pasting an instruction, to improve the instruction diversity, we introduce a mutation operation to replace the opcode with a randomly selected alternative from the instruction set (L10 and L11).
As \textit{operands} may be required by an instruction, beyond considering the operand data type, we also consider if a reference really points to an existing variable. Therefore, \texttt{fillRef} at L12 only takes the variables in $dominators$ into consideration to avoid breaking the def-use constraints of SSA form (see more details in \S\ref{sec:method:m2:prior}).
Finally, the newly defined variable and the referenced one will be recorded in $defUseMap$ (L13).
Such an iteration continues until all basic blocks are initiated.

\subsection{Liveness-Guided Instruction Refinement}
\label{sec:method:m2}
While \S\ref{sec:method:m1} ensures syntactic validity, syntactic validity alone does not equate to testing utility.
A syntactically correct but simple program often fails to trigger deep-seated bugs in the compiling pipeline.
To address this, our goal is to transform the valid skeleton into a test case with high testing utility characterized by deep dependency chains and tight inter-block coupling.
We achieve this by integrating three \textit{liveness-guided strategies} into the refinement process:
(1) enforcing global data flow complexity via block parameter coupling (\S\ref{sec:method:m2:merge});
(2) constructing deep Def-Use chains via priority-based operand selection (\S\ref{sec:method:m2:prior});
and (3) anchoring dependency chains to observable behaviors via sink synthesis (\S\ref{sec:method:sink}).

\subsubsection{Liveness-Aware Control Flow Merge.} 
\label{sec:method:m2:merge}
\begin{figure}[t] 
    \centering 
    \includegraphics[width=0.5\columnwidth]{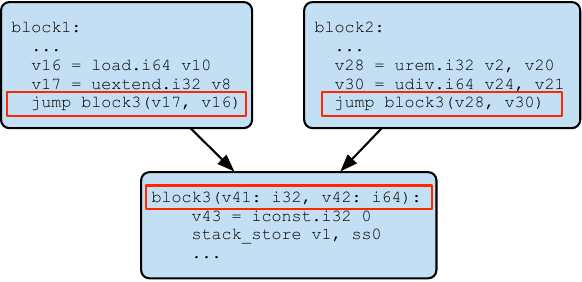} 
    
    \caption{Constructing block parameters for control flow merge points.}        
    
    \label{fig:block params}
\end{figure}

In Cranelift IR, control flow merge points are handled using \textit{block parameters} rather than Phi nodes, presenting a unique opportunity to enforce liveness across basic block boundaries.
If a merge block (\textit{i.e.,} \texttt{block3} in Fig.~\ref{fig:block params}) defines no parameters, the variables computed in its predecessors (\texttt{block1}, \texttt{block2}) are likely to be identified as unused at the end of their respective blocks, making them prime targets for optimization.

To prevent this, we employ a type-matching strategy to construct block parameters, as illustrated in Fig.~\ref{fig:block params}. 
When configuring a merge block, we analyze the \textit{live-out} variable sets of all its predecessors. 
For instance, both \texttt{block1} and \texttt{block2} possess available \texttt{i32} and \texttt{i64} variables (\texttt{v17} and \texttt{v16} in \texttt{block1} and \texttt{v28} and \texttt{v30} in \texttt{block3}). Based on this availability, we define \texttt{block3} to accept parameters of these types (\texttt{v41: i32, v42: i64}). 
Consequently, we populate the terminal jump instructions of the predecessors with specific arguments: \texttt{block1} passes \texttt{(v17, v16)} and \texttt{block2} passes \texttt{(v28, v30)}.
This explicitly binds the computations in the predecessor blocks to the execution of the successor, forcing the compiler to retain the instruction chains that produced these values.

\subsubsection{Priority-Based Operand Selection.} 
\label{sec:method:m2:prior}
Within individual basic blocks, preventing generated instructions from being optimized away requires ensuring that their defined variables are consumed by subsequent operations. To this end, when choosing operands for new instructions (L12 in Algorithm~\ref{algorithm:function generation}), we consider the following three variable categories.

\begin{itemize}[leftmargin=*]
    \item \textbf{Function Call Results.} Instructions such as function calls are computationally expensive and critical to test. Thus, to prevent ``dead call sites'', we intentionally keep a pool for function call results to prevent a function from being called but its results from being ignored.
    \item \textbf{Block Parameters.} In blocks that accept parameters (like \texttt{block3} in Fig.~\ref{fig:block params}), we maintain a pool for such parameters, which ensures that they are not merely syntactic placeholders but functional components of the block's logic.
    \item \textbf{Defined Variables.} Except for the above two categories, we also take ordinarily defined variables into consideration, like \texttt{v6} in \texttt{v6 = iadd v0, v5}. This mimics realistic programming patterns and reduces the window for a variable to be considered ``dead'' between its definition and usage.
\end{itemize}

When an instruction requires one or more operands, the generator randomly selects one of the above pools and then chooses \textit{the most recently defined variable} whose type matches the required operand type. Specifically, among all type-compatible candidates in the selected pool, we select the one with the largest ID, as it is typically the one that has just been defined. 
This strategy keeps the def-use distance short and forms a tighter def-use chain within the basic block, thereby reducing the chance that the generated instruction is regarded as dead code and eliminated by compiler optimizations. 
This operand selection process is repeated until all variable operands required by the current instruction have been filled.

\subsubsection{Sink-Anchoring for Transitive Liveness}
\label{sec:method:sink}
Improving connectivity among variables in blocks should also be guaranteed by not discarding the final results.
Therefore, in the final pass, we identify all leaf nodes of the dependency graph, \textit{i.e.,} variables that are defined but never used, or used only in non-escaping computations.
At the function's exit blocks, we deliberately synthesize strictly typed \texttt{return} instructions or memory \texttt{store} operations that consume these leaf variables.
This explicitly anchors the refined dependency chains to observable program behaviors, forcing the compiler to preserve the entire upstream computation logic.

\subsection{Diagnosis-Guided Cross-Architecture Adaptation}
\label{sec:method:m3}
As highlighted in \textbf{Challenge \#3}, differential testing of compilers faces significant hurdles due to architectural diversity. Variations in supported instruction sets and calling conventions make it difficult to utilize a single IR test case comprehensively across multiple backends. 
Furthermore, in large-scale testing campaigns, efficiency is often hampered by redundant inconsistencies triggered by the same underlying root cause. To address these challenges, we propose a \textit{diagnosis-guided cross-architecture adaptation} strategy that dynamically tailors generated IR test cases according to target environments while actively filtering out known issues.

\begin{figure}[t] 
    \centering 
    \includegraphics[width=0.9\columnwidth]{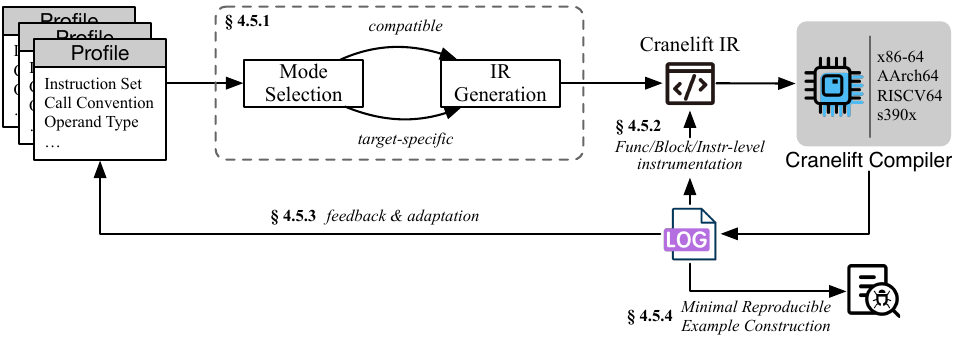} 
    \vspace{-0.1in}
    \caption{The workflow of the diagnosis-guided cross-architecture adaptation strategy.} 
    \label{fig:differential testing} 
\end{figure}

\subsubsection{Dual-Mode Testing Configuration.} 
\label{sec:approach:testing:clustering}
To balance cross-platform compatibility with architecture-specific depth, we design different profiles for different architectures. Each profile explicitly defines the supported instruction set, operand type constraints, and calling conventions for a specific backend. As shown in Fig.~\ref{fig:differential testing}, this strategy operates in two distinct modes.
In \textit{compatible mode}, the generator computes the \textit{intersection} of features supported by all active target profiles. This ensures that the generated IR test cases are portable and syntactically valid across all backends, enabling broad differential testing.
In contrast, \textit{target-specific mode} enables the generator to fully exploit the capabilities of a single backend by generating test cases tailored to its complete feature set (\textit{e.g.,} using some specific SIMD instructions only available on x86-64). This configuration improves both the breadth of cross-architecture testing and the depth of intra-architecture feature coverage.

\subsubsection{Instrumentation-Based Root Cause Diagnosis.}
\label{sec:approach:testing:locating} 
A single bug often manifests across thousands of generated test cases, resulting in a deluge of failure reports. Manually analyzing these reports is impractical. 
To address this, we propose a two-phase diagnosis method designed to assist in root cause localization.

\textit{\textbf{Phase I: Signature-guided Failure Clustering.}}
To systematically manage the volume of inconsistencies, we employ a clustering strategy based on failure symptoms. We define the \textit{failure behavioral signature} $\mathcal{S}$ as $\langle \mathcal{R}, \mathcal{K} \rangle$, where $\mathcal{R}$ represents the process exit status (\textit{e.g.,} specific crash signals) and $\mathcal{K}$ denotes a canonical set of diagnostic patterns extracted from the error stream (\textit{e.g.,} panic strings or assertion failures). 
Cases sharing an identical $\mathcal{S}$ are grouped into a single cluster, under the assumption that they stem from the same root cause.
Within each cluster, instead of random selection, we prioritize the test case with the minimum IR instruction count and the simplest control flow density. We designate this most concise instance as the cluster's representative, which significantly reduces the analysis effort required for subsequent root cause diagnosis.

\textit{\textbf{Phase II: Hierarchical Fault Localization.}}
For the identified representative, we devise a \textit{top-down fault localization} technique based on static instrumentation, \textit{i.e.,} narrowing the search scope from the coarse-grained function level down to the fine-grained instruction level. 
Specifically, we perform the localization in the following two distinct stages:

\begin{itemize}[leftmargin=*]
    \item \textbf{Coarse-grained Localization (Function \& Block).} 
    To efficiently narrow down the fault scope, we employ a top-down pruning strategy utilizing execution barriers (\textit{e.g.,} early-return instructions).
    The core mechanism involves generating program variants by inserting barriers at specific execution points: immediately following function call sites (for function-level localization) and at the tail of basic blocks (for block-level localization).
    This allows us to truncate execution and isolate the fault via differential observation.
    Specifically, if the inconsistency remains observable despite the early return inserted after a call or block, the fault must have occurred within the executed path (\textit{i.e.,} inside the called function or the current block sequence).
    In contrast, if the inconsistency disappears, the fault lies in the code subsequent to the barrier.
    We apply this logic hierarchically: first identifying the specific function containing the bug, and then recursively pinpointing the exact basic block within that function's CFG.

    \item \textbf{Instruction-level Localization.} 
    Finally, to identify the responsible instruction within the suspect block, we propose an \textit{SSA-preserving instruction localization} strategy. 
    Direct deletion is infeasible in Cranelift IR as it breaks Def-Use chains mandated by the SSA form. 
    Instead, we employ a semantic substitution mechanism: we replace a suspect arithmetic or logical instruction (\textit{e.g.,} \texttt{v2 = iadd v0, v1}) with a simplified constant assignment (\textit{e.g.,} \texttt{v2 = iconst.i32 1}). 
    If the inconsistency disappears after this substitution, the original instruction is identified as the root cause; otherwise, it is deemed irrelevant and pruned from the search space.
\end{itemize}

\subsubsection{Feedback-driven Profile Adaptation.}
\label{sec:approach:testing:adaptation}
Feedback-driven profile adaptation helps expose bugs hidden behind previously diagnosed failures and reduces redundant generation cycles. During test case clustering, a single cluster grouped by its primary failure signature may contain multiple test cases with different instruction patterns, each triggering distinct underlying bugs. For instance, a non-crashing cluster whose outputs diverge across architectures might group multiple test cases that independently trigger different backend bugs but manifest the same cross-architecture divergence symptom. Without intervention, these distinct bugs can remain hidden beneath the common failure signature. Additionally, in large-scale testing campaigns, repeatedly rediscovering known bugs wastes valuable generation cycles.

The mechanism operates through an iterative feedback loop.
After Phase II hierarchical localization identifies a root cause, the framework extracts a failure signature $\mathcal{S}_{\text{fail}}$ (typically represented as $\{\textit{Opcode}, \textit{Type}\}$). To avoid accidentally modifying the wrong instruction pattern, we manually mask this signature in the generation profile and then re-run the entire testing pipeline with the updated constraints. If the same failure cluster persists, it suggests that another underlying bug may still exist in that cluster, so the diagnosis process repeats.
Although this iterative process requires minor manual effort on updating profiles and rerunning the testing pipeline, it progressively uncovers hidden bugs and reduces human labor compared to exhaustively triaging all individual test cases. Concretely, when diagnosis identifies that a specific instruction-type combination, such as $\texttt{vhigh\_bits}$ on type $\texttt{f32x4}$, triggers a crash on a specific backend, the framework disables this combination from its generation profile for subsequent runs.

\subsubsection{Minimal Reproducible Example Construction.}
\label{sec:approach:testing:mre}
Minimal Reproducible Example (MRE) construction translates diagnostic results into actionable bug reports through targeted manual extraction. While Phase II isolates the suspected buggy code region, human intervention remains necessary to synthesize the final test case. Since automated diagnosis already narrows the search scope, this manual reduction becomes a focused extraction rather than a time-consuming debugging process. We start from the original uninstrumented IR, extract the identified buggy snippet, encapsulate it within a minimal yet syntactically valid function context, and iteratively remove surrounding instructions that are irrelevant to the bug manifestation. This process keeps the reported issue concise and readily analyzable by compiler developers.

%% file: sec-5-implement-and-evaluation.tex
\section{Evaluation}
\label{sec:imple-eval}
In this section, we quantitatively evaluate {\framework} from various perspectives.

\noindent
\textbf{Baselines.}
\label{sec:evaluation:baseline}
As Cranelift IR can be compiled from Rust and WebAssembly, we select four state-of-the-art baselines.
Specifically, \textit{cranelift-fuzzgen}~\cite{fuzzgen} is Cranelift’s official IR test case generator, based on AFL, a coverage-guided fuzzing framework that explores diverse execution paths by mutating inputs. 
\textit{RustSmith}~\cite{rustsmith} is a fuzzer originally designed for the Rust compiler. It employs a structure-aware generation strategy to produce valid Rust programs. 
As for \textit{wasm-smith}~\cite{wasm-smith} and \textit{WASMaker}~\cite{wasmaker}, they are end-to-end WebAssembly test case generators. wasm-smith, developed by the Bytecode Alliance, as part of the official WebAssembly suite, generates random but valid WebAssembly binaries. WASMaker synthesizes WebAssembly programs by combining code snippets extracted from real-world applications.

\noindent
\textbf{Research Questions.}
\label{sec:evaluation:rqs}
Our evaluation is structured around the following research questions (RQs):

\begin{itemize}
    \item[\textbf{RQ1}:] How is the performance of {\framework} compared with baselines?
    \item[\textbf{RQ2}:] Are generated test cases complex enough compared with the baselines?
    \item[\textbf{RQ3}:] What about the contribution of components of {\framework}? 
    \item[\textbf{RQ4}:] What are the characteristics, distribution, and impact of the bugs detected by {\framework}?
\end{itemize}

For RQ1, we conduct a comprehensive evaluation of the tools' effectiveness in terms of (1) the bug-finding capability and (2) the code coverage achieved on different Cranelift backends.

For RQ2, to evaluate the complexity of generated test cases, we perform a quantitative evaluation on four metrics, \textit{i.e.,} cyclomatic complexity and dominator tree depth for \textit{structural complexity}, and def-use chain depth and instruction diversity for \textit{data complexity}, on generated test cases by {\framework} and baselines.

For RQ3, we conduct ablation studies with two variants of {\framework} to evaluate the contribution of each component: 
(1) $\framework_{\text{w/o-CFG}}$ (see \S\ref{sec:method:m1}), which disables the function and control flow construction processes, restricting each test case to a single function containing a single basic block; 
and (2) $\framework_{\text{w/o-Live}}$ (see \S\ref{sec:method:m2}), which excludes the liveness-guided instruction refinement strategies.

For RQ4, we focused on a set of representative and impactful bugs to demonstrate the effectiveness of our framework. These bugs were selected based on their severity and impact on Cranelift’s functionality.

\noindent
\textbf{Implementation.}
We implemented {\framework} from the ground up in more than 7,500 lines of Rust code.
To populate the corpus described in \S\ref{sec:method:corpus}, we selected the top 100 Rust libraries from crates.io~\cite{crates} by download count and used WasmBench~\cite{wasmbench_github}, a dataset of real-world WebAssembly binaries covering a wide range of application domains. We compiled these programs into Cranelift IR, extracted their basic blocks, and serialized the blocks in a structured JSON-based format for efficient retrieval.
Based on pilot experiments comparing several mutation rates on Cranelift IR, we set the mutation rate in the test generation configuration to 50\%, which empirically balances semantic exploration with the preservation of structurally valid IR.

\noindent
\textbf{Experimental Setup.}
\label{sec:evaluation:setup}
All experiments were conducted on a dedicated server running Ubuntu 22.04, equipped with a 64-core AMD EPYC 7713 processor and 256 GB of RAM.
We evaluated {\framework} under five testing scenarios based on the \textit{Dual-Mode Testing Configuration} (\S\ref{sec:approach:testing:clustering}): one compatible-mode scenario that uses intersection-based IR for differential testing across all backends, and four target-specific scenarios tailored to x86-64, AArch64, s390x, and RISCV64, respectively. This setup allows us to exercise both cross-backend behaviors and backend-specific constraints.
To expose compiler behaviors under different optimization paths, we applied three optimization levels (\texttt{none}, \texttt{speed}, and \texttt{speed\_and\_size}) to every generated test case.
For coverage measurement, we focused on the IR-to-machine-code lowering components, \textit{i.e.,} code in \texttt{isle\_x64.rs}, \texttt{isle\_aarch64.rs}, \texttt{isle\_s390x.rs}, and \texttt{isle\_riscv64.rs}. We measured coverage hourly using cargo-llvm-cov~\cite{llvmcov} and restricted this measurement to the first 24 hours, since coverage tends to saturate within this period.
The main fuzzing campaign was executed continuously for 72 hours across all configurations. To ensure the statistical validity of the code coverage results, we repeated the 24-hour coverage experiments 10 times independently and used these runs to calculate the standard deviation.
For the bug oracle, we treat cross-architectural execution divergences as true positives by design, meaning any architecture-specific inconsistencies are actively flagged as potential bugs.

\subsection{RQ1: Effectiveness}
\label{sec:evaluation:rq1}
To evaluate the effectiveness of  {\framework}, we focus on how many unique compiler bugs are identified; and how thoroughly the generated inputs exercise the Cranelift backends.

\subsubsection{Bug-finding Capability.} Table~\ref{table:rq1:bug} presents the number of unique bugs detected by each tool on different Cranelift backends. 
As we can see, {\framework} covers the testing gaps left by existing SOTA baselines. {\framework} identified a total of 24 unique bugs. This is 8$\times$ more than cranelift-fuzzgen and WASMaker (both found three bugs), and 24$\times$ more than wasm-smith (only one bug). 
More specifically, the official test suite, \textit{i.e.,} cranelift-fuzzgen, does not detect any bugs on the x86-64 and s390x backends. This absence is notable for x86-64, the most mature target, which suggests that \textit{the official test suite may be saturated on well-tested paths}. In comparison, {\framework} identifies five unique bugs on x86-64 and two on s390x. 
It is also worth noting that RustSmith detects zero bugs in all architectures, \textit{indicating the difficulty of finding backend bugs through generating valid high-level source code}. Similarly, the WebAssembly-based baselines are limited to specific architectures, where wasm-smith only finds issues on x86-64, and WASMaker only on RISCV64. 
{\framework} is the only tool that exposed unique bugs across all four supported architectures.
Moreover, {\framework}, as the only one out of five tools, uncovers four bugs under the ``other'' category, \textit{i.e.,} bugs found in platform-independent components, such as the IR optimizer or interpreter. This demonstrates that {\framework} is effective for the entire compiler pipeline, not just specific code generation backends.

Regarding bug overlaps, {\framework} demonstrated comprehensive bug-finding capabilities by fully covering the effective search space of all baselines. 
Specifically, {\framework} successfully detected all three unique bugs identified by the official test suite, \texttt{cranelift-fuzzgen}. 
Furthermore, it also captured all four bugs exposed by the WebAssembly-based baselines (\textit{i.e.,} \texttt{wasm-smith} and \texttt{WASMaker}), whereas \texttt{RustSmith} failed to detect any. 
The fact that {\framework} not only encompasses all bugs found by state-of-the-art baselines but also uncovers a significant number of additional unique issues demonstrates its practical advantage in exercising Cranelift-specific backend behaviors. 
Collectively, these results underscore the critical necessity of a dedicated, Cranelift-native testing framework that can explore the deep backend states unreachable by existing tools.

\begin{table}[t]
    \centering
    \caption{Unique bugs detected and average code coverage achieved by {\framework} and baselines.}
    \label{table:rq1:bug}
        \resizebox{0.8\textwidth}{!}{%
        \begin{tabular}{@{}llccccc@{}}
            \toprule
            \multicolumn{2}{c}{} & \textbf{{\framework}} & \textbf{cranelift-fuzzgen} & \textbf{RustSmith}  & \textbf{wasm-smith} & \textbf{WASMaker} \\ 
            \midrule
            
            \multirow{6}{*}{\makecell[l]{\textbf{Unique Bug}}} 
            & \textbf{x86-64}  & 5  & 0 & 0 & 1 & 0 \\
            & \textbf{AArch64} & 5  & 2 & 0 & 0 & 0 \\
            & \textbf{RISCV64} & 8  & 1 & 0 & 0 & 3 \\
            & \textbf{s390x}   & 2  & 0 & 0 & 0 & 0 \\ 
            & \textbf{Other}   & 4  & 0 & 0 & 0 & 0 \\ 
            \cmidrule(l){2-7} 
            & \textbf{Total} & \textbf{24} & \textbf{3} & \textbf{0} & \textbf{1} & \textbf{3} \\
            
            \midrule 
            
            \multicolumn{2}{c}{\textbf{Avg. Coverage}} & \textbf{75.1\% $\pm$ 1.7\%} & 65.1\% $\pm$ 1.8\% & 53.4\% $\pm$ 0.3\% & 55.0\% $\pm$ 1.2\% & 53.9\% $\pm$ 2.9\% \\
            
            \bottomrule
        \end{tabular}
    }
\end{table}

\definecolor{mplblue}{HTML}{1F77B4}  
\definecolor{mplorange}{HTML}{FF7F0E}
\definecolor{mplgreen}{HTML}{2CA02C} 
\definecolor{mplred}{HTML}{D62728}   
\definecolor{mplpurple}{HTML}{9467BD}

\begin{figure*}[t]
    \centering
    \begin{subfigure}[b]{0.24\linewidth}
        \centering
        \includegraphics[width=\linewidth]{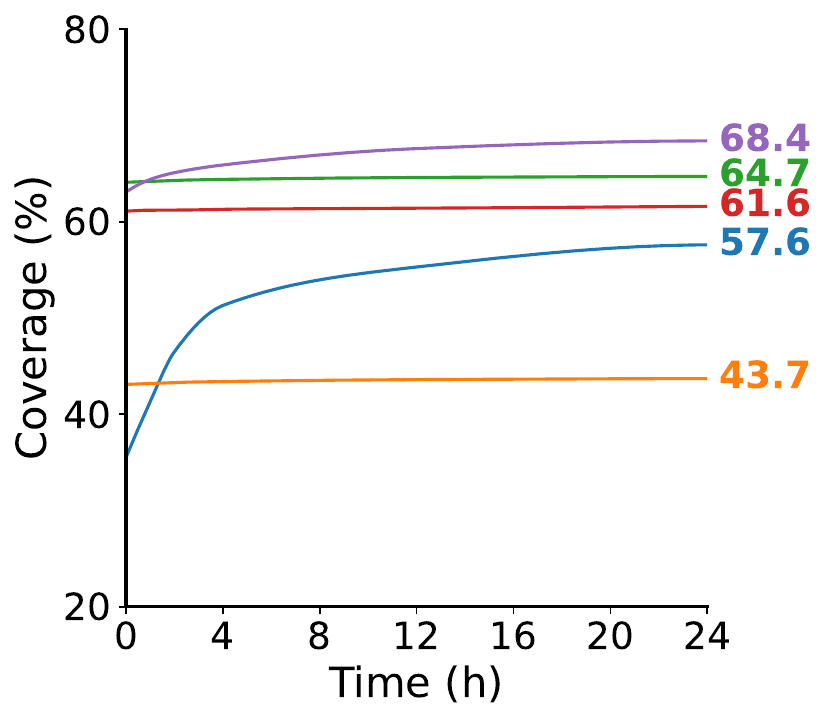} 
        \caption{x86-64}
        \label{fig:coverage-x86}
    \end{subfigure}%
    \hfill
    \begin{subfigure}[b]{0.24\linewidth}
        \centering
        \includegraphics[width=\linewidth]{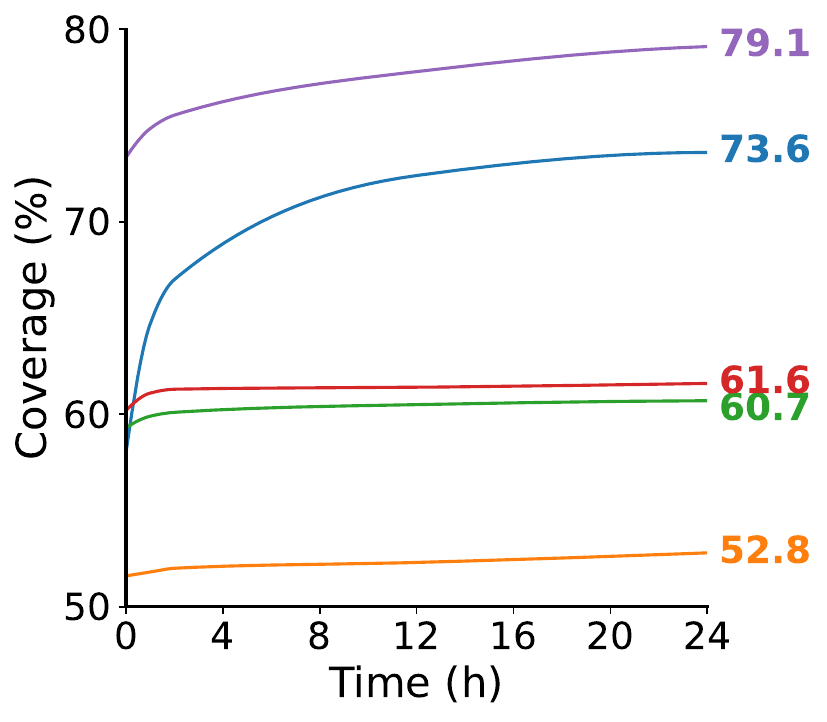}
        \caption{AArch64}
        \label{fig:coverage-aarch64}
    \end{subfigure}
    \hfill
    \begin{subfigure}[b]{0.24\linewidth}
        \centering
        \includegraphics[width=\linewidth]{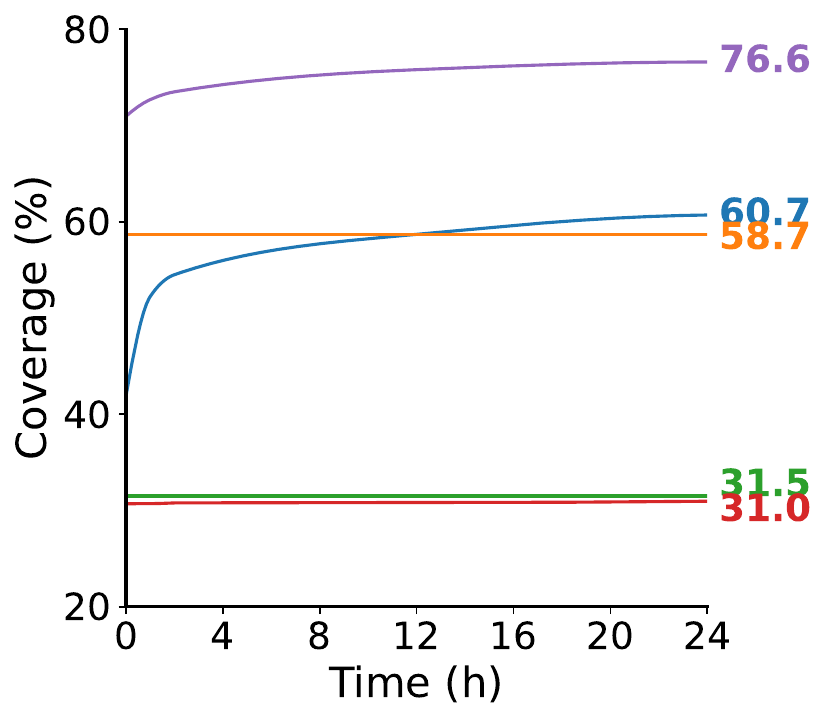}
        \caption{RISCV64}
        \label{fig:coverage-riscv}
    \end{subfigure}%
    \hfill
    \begin{subfigure}[b]{0.24\linewidth}
        \centering
        \includegraphics[width=\linewidth]{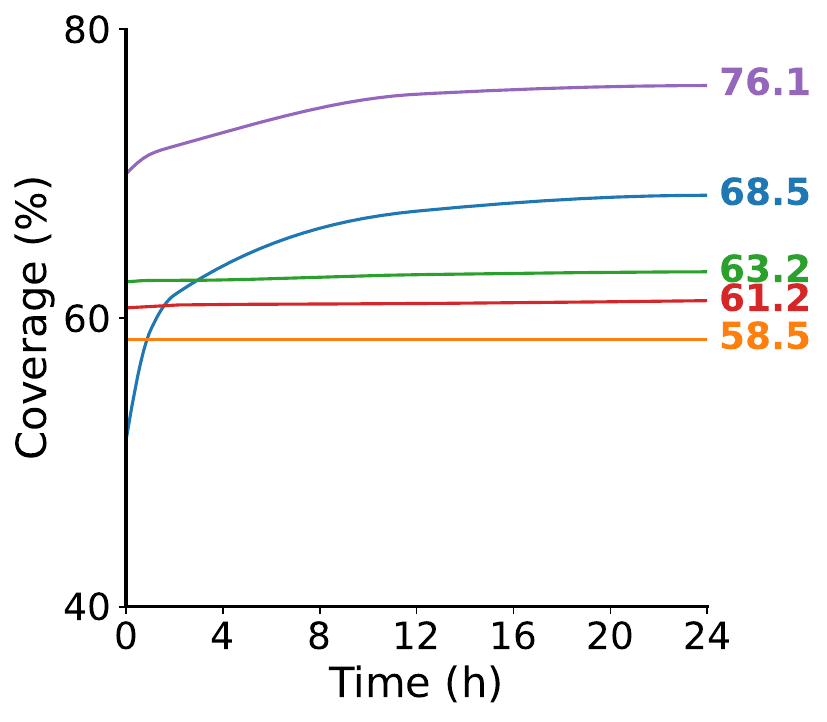}
        \caption{s390x}
        \label{fig:coverage-s390x}
    \end{subfigure}
    \vspace{-0.1in} 
    \caption{The code coverage of \textcolor{mplpurple}{\framework} and baselines on four arches. The colors correspond to \textcolor{mplblue}{cranelift-fuzzgen}, \textcolor{mplorange}{RustSmith}, \textcolor{mplgreen}{wasm-smith}, \textcolor{mplred}{WASMaker}, and \textcolor{mplpurple}{\framework}.}
    \vspace{-0.1in} 
    \label{fig:code_coverage_all}
\end{figure*}

\subsubsection{Code Coverage.} 
Code coverage reflects how much of the compiler’s code is exercised during testing. Typically, the higher the coverage, the more effective the testing. 
Since Cranelift is a backend compiler, we focus on the codebase responsible for lowering Cranelift IR to machine instructions, as mentioned in \S\ref{sec:evaluation:setup}.

As shown in Fig.~\ref{fig:code_coverage_all}, WASMaker, wasm-smith, and RustSmith quickly reach their saturation points on most Cranelift backends. This is because they are less sensitive to feedback-driven coverage expansion and can generate a large volume of test cases in a short period.
However, wasm-smith and WASMaker achieve only around 30\% coverage on RISCV64, far lower than on other architectures. After investigation, we found that when their WebAssembly binaries are compiled into Cranelift by Wasmtime, Wasmtime cannot fully utilize the architecture-specific lowering rules (as noted in \S\ref{sec:intro}). This leaves many RISCV64-specific lowering rules uncovered. This pattern also appears in other high-level language-based methods, such as RustSmith's weak performance on x86-64 and AArch64.
In contrast, cranelift-fuzzgen shows a slower but steady increase in coverage over time, because its AFL-based strategy allows each test case to gradually contribute new coverage.
{\framework} consistently starts with higher initial coverage than the baselines, demonstrating that our design explores the compiler pipeline more effectively in a shorter amount of time.
Ultimately, {\framework} achieves coverage rates of 68.4\%, 79.1\%, 76.6\%, and 76.1\% for x86-64, AArch64, RISCV64, and s390x, respectively, consistently outperforming all four baselines.
Furthermore, to evaluate overall stability, we calculated the coverage standard deviation across 10 independent runs, as shown in Table~\ref{table:rq1:bug}. {\framework} demonstrates excellent stability with a remarkably low variance of $\pm$ 1.7\%, guaranteeing reproducible exploration of deep backend paths. While RustSmith exhibits a lower variance ($\pm$ 0.3\%) due to its limited exploration space, {\framework} maintains superior stability compared to dynamically generating baselines like WASMaker ($\pm$ 2.9\%) and cranelift-fuzzgen ($\pm$ 1.8\%).

\vspace{0.1in} 
\begin{mdframed}[style=mygraybox] 
\textit{\textbf{RQ1 Answer}: Compared to current representative test schemes that can generate Cranelift IRs, {\framework} demonstrates a significant advantage in effectiveness with discovering at least 8x more unique bugs and improving coverage by at least 15\% on average.} 
\end{mdframed}

\subsection{RQ2: Structural and Computational Complexity of Test Cases}
\label{sec:evaluation:new-rq2}
To evaluate the complexity of generated test cases, we consider the \textit{structural complexity} (\textit{i.e.,} cyclomatic complexity~\cite{cc} and dominator tree depth) and \textit{data complexity} (\textit{i.e.,} def-use chain depth and instruction diversity.)

\subsubsection{Structural Complexity.}
Cyclomatic complexity characterizes the horizontal complexity and branching density, calculated as $M = E - N + P$ (where $E$ is edges, $N$ is basic blocks, and $P$ is connected components), while dominator tree depth measures the vertical depth reflecting the hierarchy of control dependencies. 
We illustrate the distribution of cyclomatic complexity and dominator tree depth for all baselines and {\framework} with different $D$ (specifying the maximum recursion depth of the CFG substitution process (detailed in \S\ref{sec:method:generation:CFG})) in Fig.~\ref{fig:semantics}(a) and (b), respectively.
As we can observe, there is a clear positive correlation between the cyclomatic complexity and $D$. When $D$ increases from 2 to 10, the generated code exhibits progressively higher logical intricacy, with the median cyclomatic complexity rising from 15.2 to 49.8. Consequently, {\framework} (particularly at $D=10$) demonstrates a significantly higher cyclomatic complexity compared to all baselines, especially the WebAssembly-based WASMaker and wasm-smith. 
Regarding the dominator tree depth, as shown in Fig.~\ref{fig:semantics}(b), {\framework} exceeds three of the four baselines.

We observe two abnormal distributions in Fig.~\ref{fig:semantics}(b), \textit{i.e.,} the one of cranelift-fuzzgen and RustSmith.
Regarding cranelift-fuzzgen, the distribution reflects its bias towards generating unit-test-like, single-block regression tests. Its generator favors straight-line code for instruction verification, leading to lower average dominator tree depth.
As for RustSmith, this anomaly is identified as a structural artifact when combined with its relatively low cyclomatic complexity (10.3). This combination reveals that RustSmith tends to generate a massive \texttt{main} function consisting of long, linear sequences of basic blocks rather than complex loop or branching logic. 
In such linear chains, every block strictly dominates its successors, artificially inflating the tree depth.

\subsubsection{Data Complexity.}
\begin{figure*}[t] 
    \centering 
    \includegraphics[width=\linewidth]{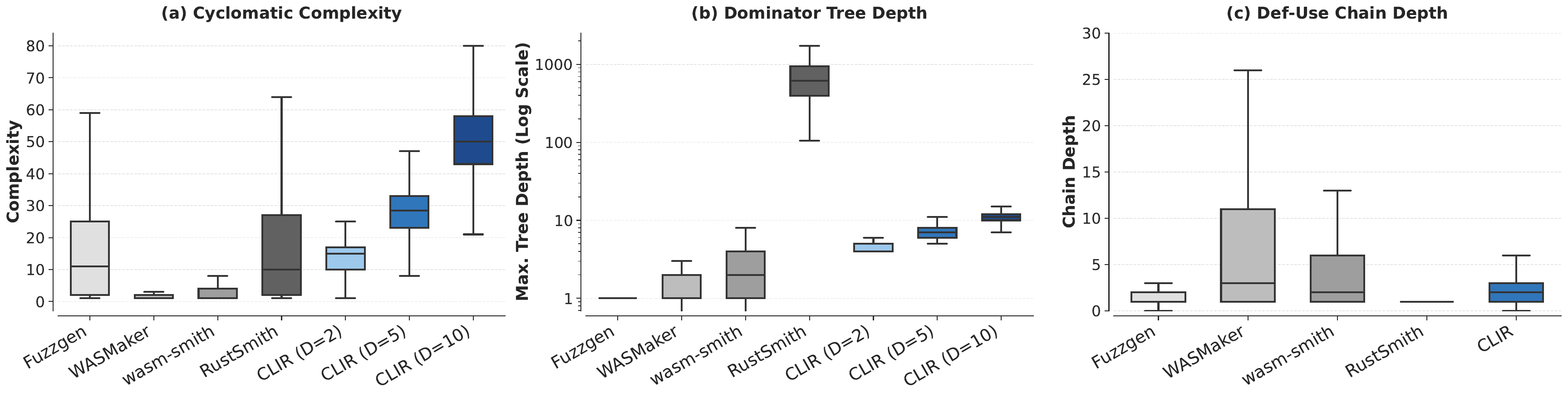} 
    \vspace{-0.2in}
    \caption{The distribution of the complexity of generated test cases in terms of (a) cyclomatic complexity, (b) dominator tree depth, (c) def-use chain depth, and (d) instruction diversity, where Fuzzgen denotes \texttt{cranelift-fuzzgen}.}
    \vspace{-0.2in}
    \label{fig:semantics} 
\end{figure*}

We take advantage of \textit{def-use chain depth} and \textit{instruction diversity} to quantitatively evaluate the data complexity.
On one hand, as illustrated in Fig.~\ref{fig:semantics}(c), both Wasm-based tools show notably long dependency chains. We find that this is due to a side effect of translating WebAssembly's stack-based instructions into Cranelift IR, where frequent push-pop operations are converted into extended, albeit often repetitive, data-flow chains.
As for RustSmith and cranelift-fuzzgen, {\framework} consistently outperforms them, achieving a median depth of 2.6 compared to $\approx$1.0 for both of them.
Finally, regarding instruction diversity, {\framework} attains a superior coverage of 89.58\%. This edges out the closest competitor, \texttt{cranelift-fuzzgen} (84.90\%), while far exceeding the other baselines (25.00\%--37.50\%).

\vspace{0.1in}
\begin{mdframed}[style=mygraybox] 
\textit{\textbf{RQ2 Answer}: The quantitative assessment concludes that {\framework} can achieve a balance between breadth and depth in terms of structural complexity, and outperforms the remaining baselines in terms of data complexity after excluding two WebAssembly-related biased ones.} 
\end{mdframed}

\subsection{RQ3: Ablation Study}
\label{sec:evaluation:rq2}

\begin{table}[t]
	\centering
	\caption{Ablation study results, where Rate refers to \#Inconsistencies / \#Test Cases.}
	\label{table:rq2:ablation}
    \vspace{-0.1in}
	\resizebox{0.8\columnwidth}{!}{%
		\begin{tabular}{@{}lccccc@{}}
			\toprule
			\textbf{}               & \textbf{\#Test Cases}   & \textbf{\#Inconsistencies}  & \textbf{Rate} & \textbf{\#Bug}    & \textbf{Avg. Coverage}      \\ \midrule
			{\framework}         & 1,327,581             & \textbf{301,849}              & \textbf{22.74\%}    & \textbf{24}      & \textbf{75.08\%}           \\
			${\framework}_{w/o-CFG}$  & \textbf{8,351,332}           & 203,732               & 2.44\%     & 13      & 60.45\%           \\
			${\framework}_{w/o-Live}$  & 1,792,921             & 256,183               & 14.29\%    & 9      & 71.74\%           \\
			\bottomrule
		\end{tabular}%
	}
    \vspace{-0.1in}
\end{table}

To evaluate the contribution of individual components, we conducted an ablation study comparing the full {\framework} against two variants: ${\framework}_{w/o-CFG}$ and ${\framework}_{w/o-Live}$, as shown in Table~\ref{table:rq2:ablation}. The results demonstrate that the full framework outperforms both variants in bug detection, despite lower throughput. Specifically, ${\framework}_{w/o-CFG}$ generated the highest volume of test cases (8.3M, 6.3$\times$ the full version) by restricting generation to single-block functions. However, this structural simplicity severely compromised effectiveness, yielding only 13 bugs (vs. 24). This highlights that mere quantity cannot compensate for the lack of structural depth required to trigger deep backend paths.
Conversely, ${\framework}_{w/o-Live}$ achieved relatively high coverage (71.74\%) yet detected the fewest bugs (9). This indicates that while the hierarchical generation provides sufficient structural complexity to explore diverse compiler paths, the generated programs lack data-flow depth. Without liveness-guided refinement, generated instructions fail to form resilient data dependencies and are actively pruned by early simplification passes.

To further analyze the 15 bugs that were detected by the full {\framework} but not by ${\framework}_{w/o-Live}$ under the same time budget, we performed a post-hoc inspection on these cases. Our goal is not to claim that ${\framework}_{w/o-Live}$ can never generate such triggers, but to explain why it is less likely to do so within the fixed ablation time. Concretely, we inspected the bug-triggering representatives generated by {\framework} and summarized three recurring liveness bottlenecks that make bug-triggering computations harder to preserve and observe.
First, \textit{lost sink anchoring} occurs when a sensitive computation does not flow into an observable sink (\texttt{return}/\texttt{store}), making it dead or behaviorally irrelevant.
Second, \textit{shallow def-use chains} make generated instructions easy to simplify away, reducing the executed code around the trigger.
Third, \textit{lost cross-region live state} occurs when dependencies are not preserved across merge points or calls, making inter-procedural triggers harder to reproduce.
Table~\ref{table:rq3:liveness} summarizes how CLIR's liveness-guided strategies address these bottlenecks, and we use Bugs \#2, \#13, and \#3 as representative examples.

\begin{table*}[t]
    \centering
    \caption{Liveness bottlenecks in the 15 bugs missed by ${\framework}_{w/o-Live}$.}
    \label{table:rq3:liveness}
    \vspace{-0.05in}
    {\small
    \resizebox{\linewidth}{!}{%
	    \begin{tabular}{@{}p{0.24\linewidth} p{0.62\linewidth} p{0.14\linewidth}@{}}
	        \toprule
		        \textbf{Liveness bottleneck} & \textbf{How CLIR mitigates it} & \textbf{Bug IDs*} \\
	        \midrule
			        \textbf{\textit{Lost sink anchoring}} & Sink synthesis anchors bug-relevant values to an observable sink (\texttt{return}/\texttt{store}), so the computations that produce these values are more likely to remain live and be exercised. & \#1, \#2, \#5, \#6, \#16, \#22, \#23 \\
	        \textbf{\textit{Shallow def-use chains}} & Priority-based operand selection encourages long transitive use-def chains, keeping more generated code live and increasing the probability that bug-triggering instructions are executed. & \#13, \#17, \#18, \#20 \\
	        \textbf{\textit{Lost cross-region live state}} & Cross-region dataflow coupling preserves dependencies across merge points and function calls through block parameters and call argument/return-value uses, improving the reproducibility of inter-procedural triggering conditions. & \#3, \#8, \#11, \#21 \\
	        \bottomrule
	        \multicolumn{3}{@{}l@{}}{* Bug IDs are defined in Table~\ref{table:bugs}.} \\
	    \end{tabular}
    }
    }
    \vspace{-0.1in}
\end{table*}

\begin{itemize}[leftmargin=*]
    \item \textbf{Bug \#2 (\textit{Lost sink anchoring}).} This miscompilation involves \texttt{scalar\_to\_vector}. In our representative test case, the value defined by \texttt{scalar\_to\_vector} is \textit{directly returned} from the function, making the bug observable as an output divergence.
    To reliably expose such a backend bug, the value produced by the sensitive instruction must affect an \textit{observable sink} (\textit{e.g.,} it is returned or stored). Otherwise, the computation is behaviorally irrelevant and may not survive to backend-specific handling.
    In ${\framework}_{w/o-Live}$, the generator does not prioritize any variable for sink connection, making it difficult to route a bug-triggering value to an observable sink within the fixed testing budget.
    In contrast, CLIR's sink synthesis (\S\ref{sec:method:sink}) explicitly connects leaf values to typed \texttt{return}/\texttt{store} operations, turning the vector construction into an observable behavior and making the bug reproducible.

 \item \textbf{Bug \#13 (\textit{Shallow def-use chains}).} This crash is triggered by \texttt{uadd\_overflow\_trap} in the interpreter. In our test case, \texttt{uadd\_overflow\_trap} appears within a def-use chain of length 6, allowing its surrounding computations to survive until the crash manifests.
    While the bug can often be reproduced with a small minimized snippet, discovering it efficiently at scale benefits from longer def-use chains, which increase the chance that the bug-triggering instruction and its context survive simplification.
    Without liveness-guided operand selection, ${\framework}_{w/o-Live}$ tends to generate shorter and more fragile chains, so many instructions become effectively irrelevant, leaving only a small executable core and reducing the probability of hitting the crash.
    CLIR's priority-based operand selection (\S\ref{sec:method:m2:prior}) strengthens intra-block dataflow by preferentially consuming recently-defined, type-matching variables (including call results and block parameters), producing longer dependency chains and improving execution density.

    \item \textbf{Bug \#3 (\textit{Lost cross-region live state}).} This miscompilation manifests as output divergence affected by an \textit{unrelated call}, \textit{i.e.,} the call is not expected to change the values that determine the function output. This indicates that backend state across a call boundary is mishandled. In our representative test case, the call site is kept semantically necessary because its return value is consumed by subsequent computations, preventing the call from being trivially removed. 
    To expose such issues, values computed before the call must remain \textit{live} across the call and be consumed afterward (\textit{e.g.,} returned or used in subsequent computations), forcing the backend to preserve/restore the relevant state.
    In ${\framework}_{w/o-Live}$, cross-region dependencies are less likely to be enforced, so pre-call values may not remain live in the post-call region, making this class of bugs harder to reproduce.
    In contrast, CLIR ensures that state-carrying values are carried across regions and become observable.

\end{itemize}

\vspace{-0.1in}
\begin{mdframed}[style=mygraybox]
\textit{\textbf{RQ3 Answer}: All proposed methods in {\framework} are important and indispensable. Disabling any of them may increase the efficiency of test case generation, but the number of discovered bugs and code coverage will be hindered.}
\end{mdframed}
\vspace{-0.1in}

\begin{table*}[t]
    \centering
    \caption{Detailed information of all identified bugs. \Circle, \LEFTcircle, and \CIRCLE\ denote "reported", "confirmed but unfixed", and "confirmed and fixed", respectively.}
    \vspace{-0.1in}
    \label{table:bugs}
    \resizebox{\textwidth}{!}{%
        \begin{tabular}{ccccl} 
            \toprule
            \textbf{ID} & \textbf{Component} & \textbf{Type} & \textbf{Status} & \textbf{Description} \\ 
            \midrule
            
            1  & s390x       & Miscompilation & \CIRCLE  & Little/big-endian flags has no effect on load instruction \\
            2  & x86-64      & Miscompilation & \CIRCLE  & Inconsistent results from scalar\_to\_vector \\
            3  & RISCV64     & Miscompilation & \CIRCLE & Unrelated calls cause inconsistent results \\
            4  & RISCV64     & Miscompilation & \Circle & Non-deterministic NaN patterns cause inconsistent outputs \\
            5  & RISCV64     & Miscompilation & \Circle & Unaligned addresses caused i16 polyfill errors \\
            6  & AArch64     & Miscompilation & \LEFTcircle & Optimizations on band instruction mask regalloc issue \\
            7  & RISCV64     & Miscompilation & \LEFTcircle & IR interpretation divergence across targets \\
            8 & RISCV64     & Miscompilation & \Circle & Calling convention mismatch on RISCV64 \\
            
            9  & AArch64     & Compiler Crash & \CIRCLE & AArch64 crashes on bitwise ops over floats \\
            10  & x86-64/AArch64 & Compiler Crash & \CIRCLE & Crash optimizing icmp with vectors \\
            11 & RISCV64     & Compiler Crash & \CIRCLE & Call instruction causes a crash on RISCV64 \\
            12 & Other & Compiler Crash & \LEFTcircle & fmin causes crash in interpret \\
            13 & Other & Compiler Crash & \CIRCLE & uadd\_overflow\_trap causes crash in interpret \\
            14 & Other & Compiler Crash & \LEFTcircle & Interpreter miscomputes bitcast result \\
            15 & Other   & Compiler Crash & \CIRCLE & Optimizer panicked when processing vector inputs \\
            16 & x86-64      & Compiler Crash & \LEFTcircle & Missing lowering rule for \texttt{vany\_true} with \texttt{i8x16} \\
            17 & x86-64      & Compiler Crash & \LEFTcircle & Missing lowering rules for \texttt{sadd\_sat}/\texttt{usub\_sat} on vectors \\
            18 & x86-64      & Compiler Crash & \LEFTcircle & Missing lowering rule for \texttt{uunarrow} with \texttt{i64x2} \\
            19 & AArch64     & Compiler Crash & \LEFTcircle & Assertion failure during emission of \texttt{band\_not} with \texttt{f64} \\
            20 & s390x       & Compiler Crash & \CIRCLE & Missing lowering rule for \texttt{bxor} with floats \\
            21 & RISCV64       & Compiler Crash & \LEFTcircle & \texttt{br\_table} crashes due to a missing \texttt{gen\_bitcast} rule \\
            22 & AArch64       & Compiler Crash & \LEFTcircle & Missing lowering rule for \texttt{vhigh\_bits} with \texttt{f32x4} \\
            23 & RISCV64       & Compiler Crash & \LEFTcircle & Missing lowering rule for \texttt{select\_spectre\_guard} with \texttt{i8x16} \\
            \bottomrule
        \end{tabular}%
    }
\vspace{-0.2in}    
\end{table*}

\subsection{RQ4: Bug Characterization}
\label{sec:evaluation:rq3}

Table~\ref{table:bugs} summarizes all 24 distinct bugs identified in RQ1\footnote{\#10 corresponds to x86-64 and AArch64.}, along with the responsible component, bug type, status, and detailed descriptions. 
As we can see, these bugs fall into two categories: 8 miscompilations, where the compiler generates incorrect machine code leading to execution divergence (\textit{e.g.,} ABI mismatches in \#8 or incorrect instruction lowering in \#2), and 16 compiler crashes, which cause the compiler to panic. The latter are mainly attributed to missing lowering rules for specific types, like SIMD vectors and floating-point instructions (\textit{e.g.,} \#16, \#17, and \#18), and assertion failures within the optimizer (\textit{e.g.,} \#15). 
These bugs span all supported backend architectures, as well as platform-independent optimization and verification phases, further demonstrating the comprehensive testing coverage of {\framework}. 
In the following, we present three representative cases as detailed case studies.

\begin{center}

\begin{minipage}{0.6\textwidth} 
\begin{lstlisting}[caption={Big-endian loads/stores mishandled.}, label={lst:endian}, frame=single, mathescape=true]
function %main() -> i64 fast {
    ss0 = explicit_slot 32
block0:
    v1 = iconst.i64 0x0011_0022_0033_0044
    stack_store v1, ss0
    v2 = stack_addr.i64 ss0
    v3 = sload32 big v2 
    return v3
}
\end{lstlisting}
\end{minipage}
\vspace{-0.1in}
\end{center}

\textbf{Case 1: Big-endian loads/stores mishandled.}
Listing~\ref{lst:endian} presents a reduced IR test case that highlights an endianness issue. L2 declares a 32-byte stack slot \texttt{ss0}. L4–L5 store a constant into \texttt{ss0}, and L6–L8 load the value using a signed 32-bit load with the big flag, indicating big-endian semantics.
On the s390x architecture, which uses big-endian memory, the result is correct. However, on little-endian architectures, the output reflects a little-endian read. This inconsistency arises because the backend fails to honor the big flag and instead defaults to the native byte order, leading to incorrect results on non-s390x targets.

\begin{center}

\begin{minipage}{0.6\textwidth} 
\begin{lstlisting}[caption={Missing vector state sync after call.}, label={lst:call}, frame=single, mathescape=true]
function %main() -> i16x8, f64x2 fast {
    sig0 = (i64) -> i64 fast
    fn0 = u1:1 sig0
    const0 = 0x00110022003300440055006600770088
block0:
    v1 = iconst.i64 0x1f96_3ea8_4eb6_5f81
    v2 = vconst.i16x8 const0
    v3 = vconst.f32x4 const0
    v4 = fvpromote_low v3
    v5 = call fn0(v1)
    v6 = iadd v1, v5
    return  v2, v4
}
\end{lstlisting}
\end{minipage}
\vspace{-0.1in}
\end{center}

\textbf{Case 2: Missing vector state sync after call.}
Listing~\ref{lst:call} shows an IR test case involving a function call that leads to inconsistent outputs on the RISCV64 architecture. The call at L10 invokes \texttt{fn0} with input $v1$ and stores the return value in $v5$. Although $v5$ is consumed by the subsequent instruction $v6 = \texttt{iadd}\ v1, v5$, the resulting value is not returned. The function \texttt{main} instead returns $v2$ and $v4$, both computed before the call.
In principle, the call should not affect the final result because the returned values are computed before the call. However, in practice, removing the call instruction leads to different outputs. This inconsistency is caused by Cranelift's mishandling of vector register state: it fails to preserve or restore registers across the call, allowing the callee to clobber values like $v2$ and $v4$, even though these values are unrelated to the call's return.

\begin{center}

\begin{minipage}{0.6\textwidth} 
\begin{lstlisting}[caption={Optimizations mask regalloc issue.}, label={lst:opt}, frame=single, mathescape=true]
function %main() -> i64, f32 fast {
block0:  
    v0 = iconst.i64 -3524126683585344751
    v1 = f32const 0x1.66e07ap-1
    v2 = band.f32 v1, v1
    return v0,v2
}
\end{lstlisting}
\end{minipage}
\vspace{-0.1in}
\end{center}

\textbf{Case 3: Optimizations mask regalloc issue.}
Listing~\ref{lst:opt} presents an IR that causes optimization-related inconsistency on AArch64. Specifically, it triggers a register allocation panic when compiled without optimizations. The issue occurs on L5, which is a bitwise operation. Although this instruction is theoretically legal and runs correctly on other architectures, it causes a crash on AArch64 due to the backend's failure to properly handle the register pressure introduced by the instruction. As a result, \texttt{regalloc2} reports an error and aborts compilation.
However, when optimizations are enabled, the issue no longer occurs. Since \texttt{band.f32 v1,v1} is semantically equivalent to $v1$, the optimizer eliminates the instruction and treats $v2$ as an alias of $v1$. This avoids triggering the faulty register allocation path and produces the expected output.

\vspace{0.1in}
\begin{mdframed}[style=mygraybox]
\textit{\textbf{RQ4 Answer}}: {\framework} effectively and efficiently uncovered 24 unique bugs of Cranelift compiler, leading to miscompilations and runtime crashes. The fact that the majority of them were confirmed in a timely manner and that their root causes are widely distributed demonstrates the influence of {\framework} in the real world.
\end{mdframed}

%% file: sec-6-discussion.tex
\section{Threat to Validity \& Discussion}
\label{sec:threat}
\noindent
\textbf{External Validity.}
Regarding the bug fix rate, it is important to note that Cranelift is a relatively new compiler backend with limited development resources compared to established frameworks like LLVM. Although the development team has acknowledged the validity of our reported bugs, immediate fixes were not feasible for all cases due to resource constraints. Instead, the developers explicitly suggested aggregating these confirmed bugs into a centralized tracking issue. This list is intended to serve as a roadmap for future open-source contributors to address. Consequently, the presence of unfixed bugs in our results reflects the project's current development stage and community collaboration strategy, rather than a lack of severity or validity of the discovered issues.

\noindent 
\textbf{Internal Validity.} 
Although {\framework} supports hierarchical bug localization down to instruction level, we acknowledge its boundary as a lightweight diagnostic aid rather than a universally precise reducer. In particular, not all inconsistencies are attributable to a single instruction: some failures emerge from interactions across multiple instructions, blocks, or loop iterations (\textit{e.g.,} optimization cascades and scheduling/regalloc coupling). In these cases, forcing instruction-level precision may be unstable or misleading. Therefore, our workflow is explicitly fallback-aware: if instruction-level isolation is not reproducible, we roll back to basic-block level; if block-level isolation is still ambiguous, we report function-level culprit regions. This multi-tiered fallback still provides actionable guidance to developers because it substantially shrinks the triage space while preserving valid, reproducible IR artifacts.

\noindent
\textbf{Discussion on Extensibility.}
Extending {\framework} to another compiler infrastructure, such as LLVM, requires target-specific engineering, but this effort is narrower than reimplementing the full system.
The main reusable parts are the Syntax-Preserving Hierarchical SSA Generation (\S\ref{sec:method:m1}) and the Liveness-Guided Refinement (\S\ref{sec:method:m2}). These components operate on SSA structure and liveness properties rather than Cranelift-specific backend rules, so they can guide test generation for other SSA-based compilers once the target IR interface is provided.
The required engineering mainly falls into two parts. (1) \textit{Code Generator Substitution and IR Mapping.} Our prototype emits instructions using Cranelift's internal code generation API. Porting requires replacing this interface with the target compiler's IR builder library, together with opcode and type mapping and compatibility checks. (2) \textit{Architecture-Specific Profiling.} As detailed in \S\ref{sec:approach:testing:clustering}, testing a new backend requires a configuration profile that defines its supported instruction set, operand type constraints, and calling conventions.
For LLVM, we estimate that porting the core generation pipeline would require approximately six working days and around 1,800 lines of code, covering common functions, control flow, and arithmetic instructions.

\noindent
\textbf{Discussion on Failure Localization and Reduction.}
Failure localization and reduction are related but distinct tasks in our setting. Reduction aims to shrink a failing input into a smaller reproducible case, while localization aims to identify the code region or instruction that is most relevant to the failure. Reduction methods such as Hierarchical Delta Debugging (HDD)~\cite{hdd} share a high-level similarity with our localization strategy because both follow a hierarchical, top-down process from coarse-grained structures to finer-grained ones. However, HDD primarily targets tree-structured inputs such as Abstract Syntax Trees (ASTs). Through recursive decomposition, it repeatedly parses and prunes the AST to find a minimal reproducible test case. In this sense, HDD is a reduction technique: its output is a smaller failing input rather than a root-cause explanation.

Our ``Phase II: Hierarchical Fault Localization'' (\S\ref{sec:method:m3}) has a different goal and operates under different constraints. Instead of deleting program fragments, our method narrows the search scope from function to block to instruction using semantic-aware IR instrumentation. Direct structural deletion is risky for SSA-form IR because it can break Def-Use chains and data dependencies required for a valid test case. Moreover, HDD is fundamentally designed for AST-like tree structures, while compiler IR often contains graph structures with cycles, such as control-flow graphs and call graphs. Our method therefore preserves SSA form and uses execution truncation and semantic substitution to identify a suspect fault region without invalidating the test case.

%% file: sec-7-related-work.tex
\section{Related Work}

\noindent \textbf{Differential testing.}
Differential testing is an effective methodology that detects bugs by comparing outputs of multiple implementations given the same inputs. It has been successfully applied to diverse targets~\cite{classming,tensorscope,compiler3,barany2018finding,wadiff}. 
Classming~\cite{classming} conducts differential testing on runtime-optimized JVMs by dynamically generating valid bytecode through real-time mutations of seed bytecode files, while Tensorscope~\cite{tensorscope} employs joint constraint analysis to generate test cases for differential testing of deep learning framework APIs, aiming to detect inconsistencies and security vulnerabilities in model conversions.
Robin Morisset et al.~\cite{compiler3} perform differential testing on compiler optimizations by comparing traces of key values before and after optimizing a given test program.

\noindent \textbf{Test case generation.}
There has been work on compiler test case generation~\cite{csmith,spe,li2024boosting,irfuzzer,hirgen,chen2019history}.
CSmith~\cite{csmith}, one of the most prevalent compiler testing tools, generates C programs through grammar-guided synthesis, producing diverse program structures that can expose bugs in compiler implementations.
Zhang et al.~\cite{spe} proposed a test case generation method called Skeletal Program Enumeration, which systematically triggers optimization bugs by exhaustively enumerating all variable usage combinations within given syntactic skeletons to generate small yet diverse test programs.
Creal~\cite{li2024boosting} constructs test cases by semantically fusing real-world code fragments, extracting functions from existing projects and combining them through dynamic analysis to achieve richer feature coverage than synthetic generation.
In addition to source-level approaches, some efforts generate IR-level test cases. 
IRFuzzer~\cite{irfuzzer} targets the LLVM backend by generating various LLVM IR programs with structured control flow and vector types. It employs constrained mutations to maintain input validity and uses instrumentation-based feedback, including matcher table coverage, to guide the fuzzing process.
HirGen~\cite{hirgen} targets bugs in the high-level IR optimization stage of deep learning compilers. It generates diverse and valid IRs by applying coverage-driven computational graph generation and leveraging high-level IR language features.
In contrast to previous work that focuses on source-level synthesis or targets specific IR stages, our {\framework} generates test cases directly in the Cranelift IR with syntactic and semantic awareness.

\section{Conclusion}
This paper presents {\framework}, a comprehensive differential testing framework designed for the Cranelift compiler. 
{\framework} employs a structure-aware hierarchical generation strategy that extracts basic blocks from real-world corpora and assembles them using a dominator-driven algorithm to ensure SSA compliance. 
To further enhance testing utility, it integrates liveness-guided instruction refinement to maximize computational complexity and employs a diagnosis-guided cross-architecture adaptation mechanism for efficient root cause localization.
Experimental evaluation demonstrates that {\framework} significantly outperforms existing techniques. 
It detects 8$\times$, 24$\times$, and 8$\times$ more unique bugs than state-of-the-art baselines (\texttt{cranelift-fuzzgen}, \texttt{wasm-smith}, and \texttt{WASMaker}), respectively, whereas \texttt{RustSmith} failed to detect any. 
Moreover, {\framework} achieves 75\% average code coverage across all supported architectures, surpassing baselines by 1.2$\times$.
Within 72 hours of testing, {\framework} discovered 24 unique bugs. 
Notably, 21 of these have been confirmed by Cranelift developers and 9 have already been fixed, demonstrating its practical impact in improving compiler reliability.

\section*{Data Availability}
The artifact of {\framework} is released at \href{https://github.com/CLIR479/CLIR}{link}.